# Large-Scale Statistical Analysis of Defect Emission in hBN: Revealing Spectral Families and Influence of Flakes Morphology

M. S. Islam[†], R. K. Chowdhury[†*], M. Barthelemy, L. Moczko, P. Hebraud, S. Berciaud, A. Barsella, and F. Fras[*]

Université de Strasbourg, CNRS, Institut de Physique et Chimie des Matériaux de Strasbourg, UMR 7504, F-67000 Strasbourg, France

**Abstract**

Quantum emitters in two-dimensional layered hexagonal boron nitride are quickly emerging as a highly promising platform for next-generation quantum technologies. However, precise identification and control of defects are key parameters to achieve the next step in their development. We conducted a comprehensive study by analyzing over 10,000 photoluminescence emission lines, revealing 11 distinct defect families within the 1.6 to 2.2 eV energy range. This challenges hypotheses of a random energy distribution. We also reported averaged defect parameters, including emission linewidths, spatial density, phonon side bands, and the Debye-Waller factors. These findings provide valuable insights to decipher the microscopic origin of emitters in hBN hosts. We also explored the influence of hBN host morphology on defect family formation, demonstrating its crucial impact. By tuning flake size and arrangement we achieve selective control of defect types while maintaining high spatial density. This offers a scalable approach to defect emission control, diverging from costly engineering methods. It highlights the importance of investigating flake morphological control to gain deeper insights into the origins of defects and to expand the spectral tailoring capabilities of defects in hBN.

***Introduction.*** Quantum emitters in two-dimensional (2D) layered materials are growing rapidly as one of the most promising building-blocks for next generation quantum technologies [1-6]. These optically active defects inherit exceptional properties from their 2D hosts, which exhibit a high mechanical flexibility, easy integration into photonic chips [7-8], and capacity to be stacked with other 2D materials (2DM) to form innovative device architectures [9-10]. Additionally, due to the atomically thin nature of their host material, the emitters are positioned in close proximity to the interfaces, enabling a high photon extraction efficiency. It also allows for the development of novel techniques for defect engineering including deterministic positioning of emitters and the ability to tune their optoelectronic properties on demand [11-

---

[†]These authors contributed equally to this work
[*]Corresponding authors: fras@ipcms.unistra.fr, chowdhury@ipcms.unistra.fr

12]. Among quantum emitters hosted in 2DM, atomic-like defects in hexagonal boron nitride (hBN) present serious advantages. They exhibit bright photostable single photon emission at room temperature and over a large wavelength range [3, 13-19], including ultraviolet [20-22] and near-infrared [23] domains. Recently, multiple studies involving resonant excitation [24], Rabi oscillation [25], ultrafast coherent control [26], and optical spin orientation [27-29] have been reported, which further strengthen hBN defects as a valuable platform for fundamental research and potential advancements in quantum technologies. However, to control the defects formation and reach the next step in their applications, it is crucial to understand the atomic composition and structure of these defects.

In the visible range, the defect emission is particularly dispersed, and the identification of its origin is the subject of an active debate. Notably, numerous theoretical works have confirmed that hBN lattice has the ability to host various type of quantum defects [30-33]. Despite these efforts, a certain number of hypotheses are still in competition, especially in the 1.6-2.2 eV optical range. Indeed, several studies suggest carbon substitutions as a key component of the defect-induced emission [34-39], but the donor-acceptor pairs could also explain the sharp luminescence lines in particular at low temperatures [40]. In addition to that, other types of hBN defects have been assigned in the visible range, such as dangling bonds appearing at surfaces or grain boundaries [41], natural vacancies or antisites [42-45], and oxygen substitution [46]. This difficulty to decipher the origin of hBN defects is fostered by the lack of precise experimental spectral assignments of the defect emission lines. The latter is rendered particularly difficult due to (i) the possible coexistence of several different defects lying in close spectral and spatial domains, (ii) the complex defect emission spectrum composed by the main transition line, labelled as zero phonon line (ZPL), frequently accompanied by one or multiple phonon side bands (PSBs), and (iii) the spectral variability induced by local defect environment such as strain or electric field from trapped charges and dielectric surroundings.

In ref. [15, 47], the authors reported histograms of ZPL spectral occurrences based on a sample size of 40–100 emission lines at room temperature. The histograms show a broad spectral distribution of ZPL occurrences from 1.6 to 2.2 eV, which indicates, at a first glance, that ZPL emission can appear continuously in this range. At cryogenic temperatures, Jungwirth et al. [13] and Dietrich et al. [48] presented ZPL histograms composed of 340 and 627 occurrences in total with spectral binning of 10 meV and 10 nm, respectively. Both histograms show similar continuous distribution of ZPL, but with four clear clusters of occurrences, opening an opportunity for optical assignments of defect emission families. These results highlight the fact that the question of the spectral randomness of defect emission in hBN host should be addressed across a very large number of defects to be statistically significant.

In this article we tackle the challenges of spectral identification of defect emission families in liquid exfoliated hBN nanoflakes samples at room temperature. Apart from the cost-effectiveness and the scaling-up possibilities offered by the liquid exfoliation process, these samples present various configuration aspects in terms of flakes size, flakes agglomeration, and local environment (Fig. 1a). Therefore, such hosts provide representative samples of the variety of defect species and their spectral distribution. We present a comprehensive investigation based over 87421 photoluminecence (PL) spectra, gathering in total 8307 ZPLs. Such sampling enables us to construct ZPL occurrence histograms with a spectral binning of 2 meV, resulting in a high resolution of the ZPL spectral distribution and then allowing us to identify narrow sets of defect families. To analyse such a high number of spectra, we developed a peak detection algorithm that determines automatically the spectral positions and linewidths of ZPLs and corresponding PSBs. As a result, 11 distinct spectral ZPL emission families are clearly identified within the 1.6 to 2.2 eV range, each associated with well-defined energy centers, resulting from a sharp discretization of the ZPL energy distribution. This challenges the hypothesis of a broad and continuous distribution of defect energy associated with spectral randomness. Such a precise identification will be valuable to figure out the microscopic origin of the emitters in hBN hosts. Indeed, the density functional theory calculations cannot usually provide the absolute transition energies with enough accuracy [32, 46]. Therefore, the spectral spacing between families is a more reliable parameter, which could be crucial in identifying the chemical composition of defects. Along with the spectral identification of defects, we also take advantage of the statistical approach to estimate significant values of the coupling constants between defects and lattice vibrations, over a large dataset. Lastly, we present a pioneering analysis exploring the role of hBN nanoflake morphologies in the formation of different defect species. The occurrences of various defect types are analyzed in relation to both the size of the nanoflake building blocks and the type of nanoflake agglomerations. Our findings demonstrate that precise control over defect family formation can be achieved by easy tuning of these morphological parameters.

***Samples, µPL experiment and peak detection scheme.*** Layered hBN solution was processed *via* sequential ultrasonication, centrifugation, and filtration steps to achieve size-selected hBN nanoflakes dispersed in a 1:1 ethanol/water mixture. Subsequently, the processed hBN solutions were deposited onto plasma cleaned $SiO_2$/Si substrates, followed by thermal annealing (Fig. 1c). Two samples were here studied (see Methods). Sample A was processed from the supernatant solution after centrifugation at 8000 rpm, and contains small flakes with sizes distributed from 80 to 160 nm (FWHM boundaries), as indicated by the dynamic light scattering (DLS) measurements (Fig. 1d). Sample B was obtained from the complementary precipitate and contains larger hBN flakes with two distinct size distributions of flakes. One spans from 130 to 250 nm, while the other extends from 9 to 14 µm (Fig. 1d). This broader distribution by two orders of magnitude implies that, in sample B, the hBN material predominantly resides within the second size distribution of larger flakes, centered around 11 µm. Next, we acquired hyperspectral maps from both the samples using home-built confocal µPL setup with diffraction-limited laser excitation at 532 nm (see Methods). With such sub-hBN-bandgap excitation, we probe the defect induced emissions ranging between 1.6 eV and 2.2 eV, at room temperature. A scheme of the setup is presented in Fig. S1 of supplementary information (SI).

As the majority of spectra contain multiple PL peaks, reliably identifying the ZPLs is particularly challenging. Within a spectrum, the different peaks can either correspond to multiple ZPLs from nearby defects or from a ZPL/PSBs spectrum originating from a single defect [16, 36, 49-50]. Recently, Hoese et al. [24] demonstrated that PSB and ZPL signatures can be identified distinctly by two photon correlation measurement. However, such analysis cannot be applied automatically to a large sample of spectra. Therefore, we have developed an algorithm which relies on the known physical properties of the defect emission spectrum in hBN hosts to identify and separate the ZPLs and PSBs. Indeed, it has been shown that mainly three hBN phonon modes lead to PSBs at room temperature [22, 24, 51-53]. They are red-shifted by approximately 166, 175, and 200 meV from the ZPL, with a significantly lower PL intensity and a larger linewidth compared to the ZPL. Therefore, to determine automatically and unequivocally whether a peak is a ZPL or a PSB, the developed algorithm proceeds as follow for each spectrum. For a single peak spectrum (SPS), it considers that peak as a ZPL. In the case of a multiple peaks spectrum (MPS), it analyses two by two the energy and prominence (relative height of a peak, see SI) differences between all detected PL peaks. If (i) the energy difference matches with the energy range of the main phonon modes and (ii) the red-shifted peak is of lower prominence and larger linewidth, the latter is not considered as a ZPL and sorted as PSB. Note also that such criteria include multi-order PSBs (two or more phonons involved). To minimize the ZPL identification errors, we adopt a non-restrictive condition on the phonon energy range (ZPL-PSB redshift) using a uniform interval of 135 to 215 meV. This allows the algorithm to tolerate small fluctuations of phonon modes due to local strain and layer thickness [54-56]. As a consequence, the ZPLs are identified unequivocally. Conversely, their number could potentially be underestimated in the case of a ZPL falling in the redshift range of another ZPL and satisfying the aforementioned criteria. A flowchart accompanied by a detailed description of the algorithm is provided in the SI.

***Identification of spectral families of defects.*** First, we present the results obtained by combining data from samples A and B, in order to form the most comprehensive statistic. To collect data from a significant number of defects, we performed 21 continuous hyperspectral maps covering 87421 different spatial positions from two different samples, gathering in total 8307 ZPLs. A typical hyperspectral PL map and representative PL spectra obtained from different map positions are presented in Fig. 1e-f. All the spectra are analysed by the algorithm, allowing us to retrieve the peak position in energy ($E_{00}$) and the corresponding linewidths ($\Delta E_{00}$). We then constructed histograms of all the ZPL peaks, with a spectral binning of 2.0 meV, resulting in the spectral distribution of defect energy. The result is presented in Fig. 1g. Remarkably, the ZPL peak spectral distribution does not appear to be continuous, but clearly separated into discrete sets. From this distribution, we identify 11 spectral families of defect emission ($F_1$-to-$F_{11}$) over the range 1.6-2.2 eV. The histogram is well fitted by the sum of 11 Gaussian functions, with a residual standard error of 22.2 and a mean occurrence of 601.5. This high accuracy suggests that nearly all of detected ZPLs belong to one of the 11 families, highlighting the relevance of assigning defect emissions to discrete families. The different family Gaussian distributions exhibit well-defined energy centers ($c_i$), characterized by widths ($w_i$) ranging from 0.2 meV to 16.8 meV (HWHM), as summarized in Table 1 (complementary data are given in SI). This observation indicates that the defects are weakly affected by the inhomogeneity of the samples such as flakes orientation, thickness, local strain, and dielectric surrounding. It is worth noting that the predominant defect families, F1 and F2 centered around 2 eV, which comprise more than 30% of the total detected defect centers, are the most commonly observed quantum emitters in hBN. On the other hand, another family, F7, exhibits a remarkably narrow dispersion of the energy center ($c_7$, $w_7$) of approximately 0.2 meV, along with a narrow mean linewidth ($<\Delta E_{00}>_7$ in Table 1) of 3.6 meV and a high occurrence, indicating an exceptional decoupling from the local environment in ambient conditions.

A large majority of spectra contain multiple PL peaks, of which typical examples are shown in Fig. 2a. Therefore, to dispel any ambiguity regarding the ZPL identification, we compare ZPL energy histograms for single peak spectra (SPS) and multi peak spectra (MPS) in Fig. 2b-c, respectively. Both histograms exhibit a similar discretization pattern in the ZPL peak energy distribution with the same sets of energy centers ($c_i$). This consistency in energy centers indicates that the spectral assignment of defect families remains independent of whether they are observed in MPS or SPS. One can note that due to a larger number of ZPLs collected from MPS, certain features become more distinguishable in the MPS histogram, leading to an easier spectral identification of defect families. However, there are some differences between the SPS and MPS distributions. Interestingly, the relative weights of the ZPL families are distinct, and particularly two of them ($F_9$ and $F_{11}$) are only visible in the MPS histogram. The area probed for each spectrum is about 1 µm². Therefore, multiple spectral signatures correspond to dense regions, where several radiative defects coexist below the micrometric scale. The presence of nearby defects in hBN were highlighted with scanning tunnelling electron microscopy in ref [57], where the authors show that multiple defects can be localized within a ∼ 35 nm region. To explain the difference between SPS and MPS ZPL distributions, we can formulate the following hypothesis. In dense defects regions, the interaction between defect centers can lead to the formation of complex or clustered defects with different electronic structures compared to isolated defects. For instance, defects with opposite charge, like donor-acceptor, can pair through Coulomb interaction [39-40, 58]. Additionally, the elastic interactions and the formation of covalent bounds can also lead to defect associations [59-61].

We also take advantage of our statistical approach to provide relevant values of the coupling between defects and lattice vibrations, collected over 2898 MPSs. Fig. 2d presents the histogram of PSB energy shift relative to the corresponding ZPL. The PSB histogram indicates three resonances, which match the energy shift of three main red-shifted PSBs mainly reported [22, 24, 51-53], which further verifies the relevance of the developed peaks sorting methodology. Additionally, in Table 1, we report key characteristics for each family such as, main PSB energies, percentage of ZPL showing a PSB, and an estimation of Debye-Waller factors (DWF). The DWF is given by the fraction of light emitted into the ZPL, and characterizes the strength of electron-phonon coupling. Along with the ZPL the values of DWF are, in return, decisive to explore theoretically the microscopic origin of defects [34, 35, 41, 43-44]. The experimental values presented in Table 1 show a significant variation of the PSB presence across different defect families, accompanied by a distribution of DWF ranging from 52.4 % to 73.7 when PSBs are detected. Overall, these results demonstrate a wide disparities of the electron-phonon coupling among the different defect families. It is important to note that, unlike the ZPL histograms, the PSB histogram may contain certain biases. In MPS, there there may be instances where two or multiple ZPLs exhibit red-shifted energies ranging from 135 to 215 meV, which could potentially lead to their misidentification as PSBs. However, the additional criterion employed in the sorting algorithm, such as linewidth and peak prominence comparison, effectively minimize these biases. We believe that the overall values presented above accurately capture the essential features of the coupling between defects and phonons.

***Agglomeration vs. defect families.*** Now that the defect families are spectrally identified, we explore the flakes morphological parameters that can control defect family formation. We first start to question the role of the flakes agglomeration size. Indeed, regardless of the liquid-exfoliated 2D material deposition processes (whether drop-casting or spin-coating), once the flakes are deposited, they tend to attach or cluster both horizontally and vertically in various ways, differing from the 3D bulk symmetry. This results in the formation of flake agglomerations of diverse sizes, and leads to a non-uniform density of hBN flakes across the sample. Therefore, comparing the contrast-profiles of white-light images and scanning electron micrographs (SEM), we classify all the PL probed areas into two flake arrangement morphologies: agglomeration size of flakes typically below a micron, which potentially results in isolated flakes (now referred to as small agglomeration), and agglomeration size of flakes above 5 µm (now referred to as large agglomeration). SEM examples of such different morphologies are shown in Fig. 3a-b. Additionally, white-light images of such characteristic PL mapped areas for both the morphologies are included in SI. Here, we consider only sample A, as it was made from a solution containing a single narrow nanoflakes size distribution centered around 100 nm. The ZPL energy histograms built for small and large agglomeration types in sample A are presented in Fig. 3C-d. The histograms exhibit clear disparities in the occurrences of defect families. Specifically, small agglomeration flakes predominantly show only two types of defects, F1 and F2 (with F7 being barely identifiable). In contrast, large agglomeration flakes display a wider range of defect families, encompassing all the identified defects except for the F3, F5, and F8 families. The physical comprehension of this trend is beyond the scope of this study; nonetheless, we can propose lines of thinking. First, flake edges can play a pivotal role in the formation of defects and their resulting optical properties. Indeed, different chemical groups can be attached to the flake edges, in particular terminations made of nitrogen dangling bonds passivated by hydrogen were identified experimentally [64]. It was also shown theoretically that for boron [43, 45, 65] and nitrogen [65] vacancies defects, such a hydrogen passivation induces a symmetry lowering and lattice distortions, leading to significant changes in the ZPL energy. The influence of edges is expected to be particularly significant in sample A, which consists of small flakes exhibiting a high density of edges. Now when individual nanoflakes agglomerate, the edges and terminations undergo modifications. The structure and symmetry of the edges could be reconfigured, and the reactive sites at the edge may become passivated or bonded to form a new termination type. After agglomeration, defects can also lie in close contact influencing each other or resulting in the formation of complex defects [59-61]. Lastly, apart from the edge effects, the presence of other layers, due to stacking, changes the electrical permittivity around the defects compared to isolated flakes. Overall, these arguments provide a plausible explanation for the wider variety of defects observed in large agglomerations compared to small ones.

***Building blocks vs. defect families.*** We now inquire about the influence of the flakes size on the occurrences of defect families. For this purpose, we conducted a comparative analysis between sample A, obtained from a solution of small flakes with a mean size of 100 nm, and sample B, made from a solution of larger flakes with two size distributions characterized by mean values of 210 nm and 11 µm, respectively (see Fig. 1d). The size-selected flakes in solutions used to prepare sample are referred to as the building blocks flakes. To highlight the impact of the size of the building blocks flakes, we specifically focused on large agglomerations. Indeed, small agglomerations (below 1µm) of sample B are excluded from consideration since they do not contain large flakes. Conversely, the large agglomerations in sample B (above 5 µm) are predominantly composed (in terms of hBN material amount) of the larger flakes centered around 11 µm, as indicated by the size distribution presented in Fig. 1d.. A comparison of ZPL occurrences between sample A and sample B is presented for large agglomerations in Fig. 4a. The host hBN made exclusively from large flakes exhibits three families (F3, F5, and F8), which are absent in the hosts made from small flakes, showcasing a distinct and complementary set of defect families between the two types of hosts. These observations provide direct evidence of the role of flake sizes in defect family formation. Here also we can formulate hypothesis to explain such observations. First, the surface-to-volume ratio decreases in large flakes compared to small ones [56]. Therefore, it can be assumed that the density of defects close to edges and surfaces is massively reduced in hBN hosts made from ~10 µm flakes size. Secondly, as discussed previously, the location of defects relative to edges can play a crucial role in their formation and optical properties. Therefore, hBN hosts composed of larger flakes are expected to exhibit deeper defects that are less influenced by the presence of edges, resulting in distinct types of defects compared to those found in agglomerations made from smaller flakes.

The latter observations prompted an exploration of the degree to which the selection of defect families can be achieved based on hBN host morphologies. As an illustration, Figure 4b presents comparative spectral histograms of ZPLs detected in small agglomerations (isolated small flakes) from sample A and large agglomerations from sample B (agglomerated big flakes). These two extreme cases reveal a clear spectral discrepancy and grouping of discernible defect families, wherein the "edge defects" from sample A ($F_1$, $F_2$) exhibit a notable blue shift compared to the "deeper defects" ($F_3$, $F_5$, $F_8$) from sample B. Additionally, it is worth noting that the 'edge defects' present the widest spectral widths distribution ($w_i$ in Table 1) compared to other defects, consistent with their atomically thin local environment potentially subjected to more fluctuation. Such a dependence of defect families on morphological flakes parameters opens up advantageous opportunities to easily control the spectral occurrences of hBN defects. By selectively choosing hBN building-block sizes and employing precise spin-coating techniques, it becomes possible to prevent or promote agglomeration, providing a straightforward approach to tailor the spectral properties of defects in hBN hosts.

***Defects density vs morphology.*** Lastly we conclude our work by an analysis of the global density of defects as a function of flake morphology, which is a crucial parameter that influences the efficiency, and functionality of potential hBN-defects-based devices. To facilitate comparison, we calculated the effective density of defects by dividing the total number of observed emitters by the area covered by hBN flakes. The details of area estimations, using the greyscale contrast profile of white-light and SEM images are presented in SI. The summary of effective defect densities for different morphologies is provided in Table 2. Remarkably, the effective defect densities (ranging from 0.26 to 0.53 $\mu m^{-2}$) are relatively similar across all cases, despite the important variations in host thickness between small agglomerations of sample A (isolated few-layers flakes) and the large agglomerations formed with large flakes of sample B. First, this suggests that the defects are predominantly hosted in the first few layers regardless of the morphologies of the hBN host. This finding confirms the hypothesis presented in Ref. [63] using a different sample preparation protocol but also involving high-temperature annealing as final step. It is worth noting that the annealing process leads to the migration of defects, potentially resulting in their formation at the surfaces and edges. Secondly, the achieved densities of defects are comparable, and even better, than one of the highest reported densities observed in mechanically exfoliated samples [62], which were treated with advanced thermal processes while preserving the material quality of the hBN host.

This highlights that morphological control of defects is compatible with achieving a high defect density.

***Conclusions.*** We have performed a comprehensive statistical study based on a very large number of hBN defects PL emission lines (more than 10000). Based on such a sampling, we identified 11 discrete spectral families of defects with well-defined defect emission centers over the spectral range from 1.6 to 2.2 eV. This finding refutes the previous hypotheses suggesting spectral randomness of defects leading to a broad and continuous distribution of their energy. We also took advantage of our statistical approach to provide relevant defect parameters such as, emission linewidths, spatial density of defects, phonon side bands, and Debye-Waller factors. Our report should be valuable to determine the microscopic origin of the emitters in hBN hosts. In particular, the spectral spacing between defect families could serve as a key parameter for theoretical investigations, rather than relying on absolute energy values, as determining the latter with sufficient accuracy is very challenging and hinders the possibility of deciphering the chemical nature of defects. Additionally, we have shown that the morphologies of the hBN host play a crucial role in the spectral occurrence of defect families. Conversely, we have demonstrated that the variety of defects can be selected by controlling the size and arrangement of the flakes, while maintaining a high density of defects. In view of applications, this aspect is particularly alluring as it can pave the way for easy and scalable control of defect emission in the hBN host, diverging from costly defect engineering methods. Finally, the newfound degree of freedom offered by flake morphology control clearly warrants further exploration in order to gain deeper insights into the underlying physics of defect formation and fully explore its potential as spectral control of defects in hBN.

**Methods**

***Sample preparation and characterization.*** Liquid exfoliated layered (1-5 layers) hexagonal boron nitride (hBN) dispersion were purchased from Graphene Supermarket. Firstly, 20 ml of the hBN dispersion was sonicated for 30 min in an ultrasonic bath to circumvent the smaller cluster formation due to sample aging. Following this, the hBN solution was centrifuged at 8000 rpm for 30 min to separate into supernatant (solution for sample A) and precipitate (solution for sample B). In this process, we achieved a size selection of the flakes in the solutions, which will serve as the building-block flakes for the hBN samples. A commercial dynamic light scattering (DLS) setup (ALV / CGS-3) was used to characterize nanoflake size distributions for both sample A and sample B in the solution phase. The acquisition is performed at a scattering angle of 90°, under illumination at 633 nm. The intensity correlation function is computed in a homodyne mode, and is then inverse Laplace transformed using a CONTIN algorithm [66]. One obtains the intensity weighted distribution of relaxation times, from which the distribution of translation diffusion coefficients and subsequently of nanoflakes hydrodynamics radii are deduced. The experiments are performed at low system concentrations, guaranteeing that the resulting distributions remain unaffected by concentration levels. Next, these hBN solutions were spin-coated onto 1:1 $O_2$/Ar plasma treated (2 min) 300 nm thick $SiO_2$/Si substrates. We found here that the controlled plasma treatment of the substrate was particularly critical to accomplish excess surface area coverage and desired morphologies. Afterwards, the as-prepared samples were annealed at 850 °C for 3 hours in a tube furnace under atmospheric pressure. It is important to mention that the large agglomerations were mainly formed during deposition and drying of nanoflakes on the substrates. All the samples were characterized by a scanning electron microscope to explore different morphologies of hBN nanoflakes distribution over the substrates.

***Optical characterization.*** A home-built confocal PL microscopic (0.6 NA, 50x objective) mapping setup along with white-light imaging was used for characterizing our fabricated hBN samples. A continuous-wave 532 nm laser was used to excite the samples. The PL emissions were collected using a spectrometer with a spectral resolution of 0.15 nm over the 555-785 nm (1.58-2.23 eV) range. The excitation power was maintained at 2 mW for all the PL measurements. This ensured saturation for the majority of emitters and allowed us to select the most robust ones. Each spectrum of the hyperspectral maps was acquired with an acquisition time of 1 sec. The samples were placed on a computer-controlled piezoelectric stage. The consecutive pixels in the hyperspectral maps were separated by a one µm step of the piezo-stages. The combination of such a step size and the confocal detection allowed for the prevention of multiple counting from localized emitters. All the optical measurements were conducted at room temperature.

**Acknowledgements**

The authors would like to thank Tom Ferte and Cedric Leuvrey for SEM imaging. This research was supported by the Agence Nationale de Recherche (ANR) under the grants FINDING ANR-18-CE30-0012-01 and LHNANOMAT ANR-19-CE09-0006, and from the QUSTEC program (European Union's Horizon 2020, Marie Skłodowska-Curie Grant Agreement No. 847471).

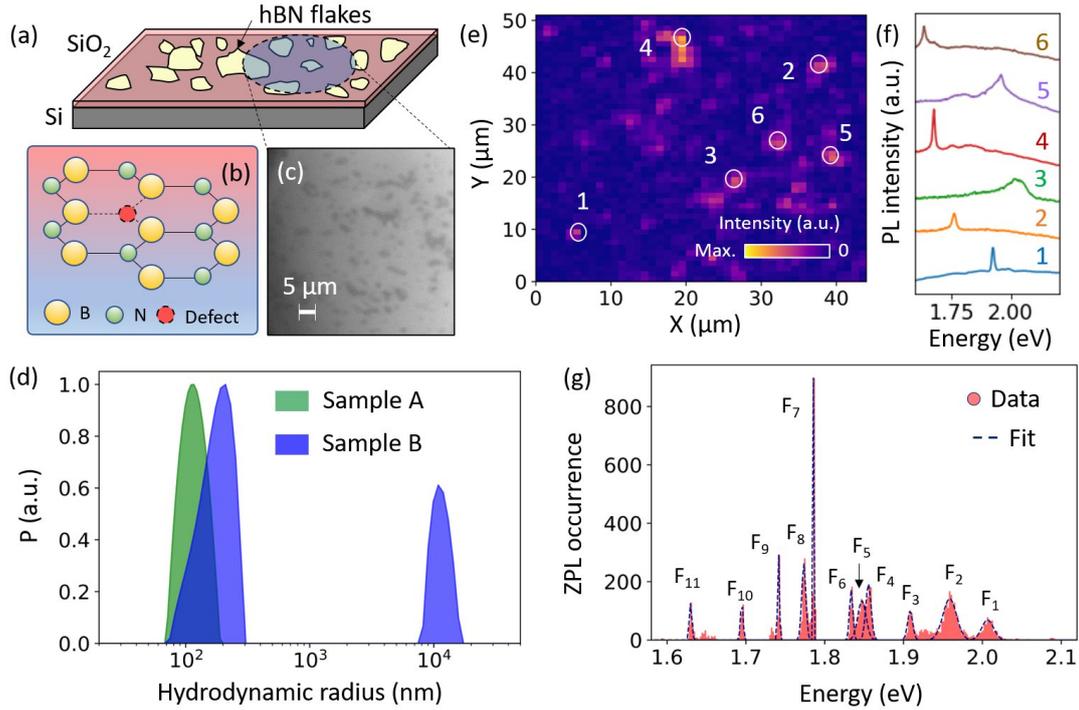

**Figure 1: Characterization of hBN samples.** Schematic representations of (a) hBN nanoflakes sample and (b) hBN lattice with a defect located on a vacancy site for example (c) White light image of a typical area on hBN sample, other examples are provided in SI (d) Intensity weighted size distribution P of hBN flakes for samples A (green) and sample B (blue) obtained from DLS measurement (e) An exemple of an hyperspectral PL map and (f) six typical single-peak PL spectra (with offset along normalized intensity) (g) Histogram (red coloured, number of total ZPLs: 8307) of total ZPLs occurences collected from both samples A and B, fitted by the sum of 11 Gaussian functions (dashed line, residual standard error of 22.2 and a mean occurrence of 601.5.), identifying 11 defect families ($F_1$-to-$F_{11}$).

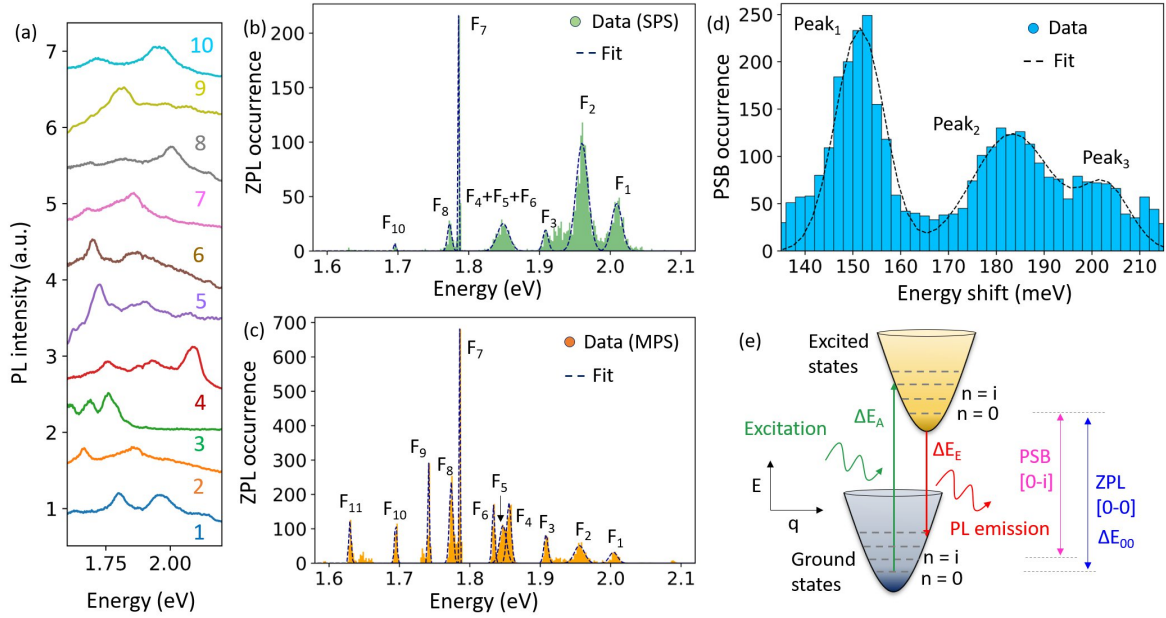

**Figure 2: Single-peak spectra (SPS) vs. multi-peak spectra (MPS).** (a) Stack-plot (with offset along normalized intensity) of 10 typical multi-peak spectra (MPS) in hBN (b) SPS with green coloured (number of ZPLs: 2564) histogram and (c) MPS with orange coloured (number of ZPL: 5743) histogram. All the defect families are identified with multi-gaussian histogram fitting (d) Histogram (cyan coloured, number of PSBs: 3501) of all detected PSBs from MPS along with multi-gaussian fitting (dashed lines), indicating three peaks (e) Schematic representation of vibrational energy levels in hBN sub-bandgap-defects along with excitation, ZPL, and PSB emission processes. Here, E as energy, q as reaction coordinate, $\Delta E_A$, $\Delta E_E$, and $\Delta E_{00}$ as excitation, emission, and ZPL energies, respectively.

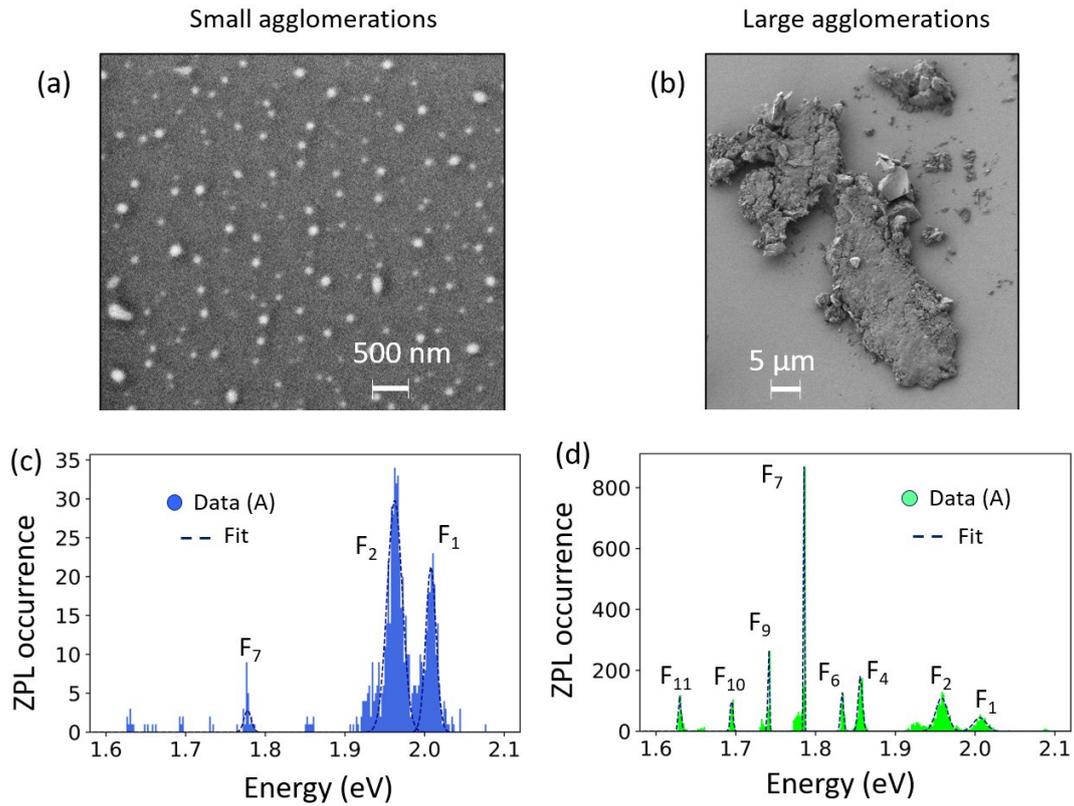

**Figure 3: Impacts of hBN agglomeration types.** Representative scanning electron microscope images of (a) small and (b) large hBN agglomerations in sample A. (c) ZPL histograms and corresponding fits (dashed line) for small hBN agglomerations (blue colored histogram, number of ZPLs: 783) in sample A. (d) ZPL histograms and corresponding fits (dashed line) for large hBN agglomerations (lime colored histogram, number of ZPLs: 4839) in sample A.

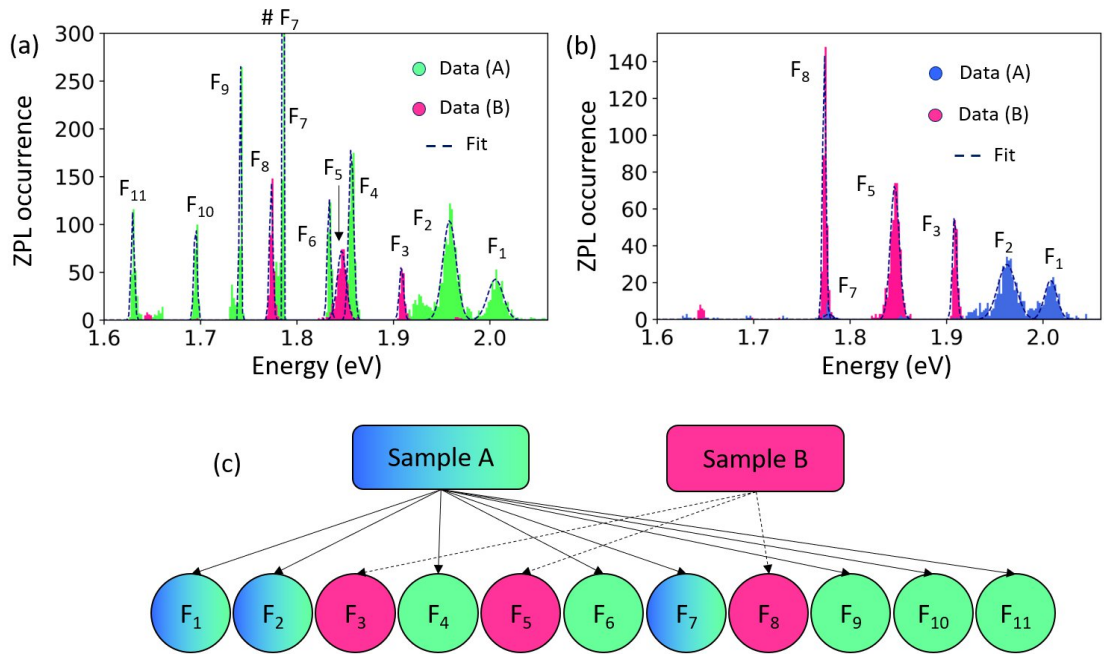

**Figure 4: Impacts of hBN building blocks.** (a) Comparative histogram profiles and corresponding fits (dashed line) as a function of large agglomeration for sample A (lime coloured histogram, number of ZPL: 4839) and sample B (pink coloured histogram, number of ZPL: 991). Value of ZPL peak maxima for $F_7$ (# $F_7$) is 846. (b) Comparative histograms and corresponding fits (dashed line) of defect families for two extremely distinct morphological cases: small hBN agglomeration from sample A (royal blue coloured histogram, number of ZPLs: 783) vs. large hBN agglomeration from sample B (deep pink coloured histogram, number of ZPLs: 991) (c) Schematic summary showcasing the presence of different defect families for samples A and B.

| Defect family | Center $c_i$ and width $w_i$ (HWHM) of ZPL peak energy distributions | | Mean ZPL linewidth $<\Delta E_{00}>_i$ (HWHM) (meV) | Relative abundance (%) with respect to total detected ZPL | | Mean PSB energy (eV) | Proportion of ZPL with PSB (%) | Mean DWF (when PSB detected) |
|---|---|---|---|---|---|---|---|---|
| | $c_i$ (eV) | $w_i$ (meV) | | Sample A | Sample B | | | |
| $F_1$ | 2.01 | 14.8 | 10.2 | 10.7 | None | 1.85 | 22.4 | 65.3 |
| $F_2$ | 1.96 | 16.8 | 11.3 | 20.2 | 0.4 | 1.78 | 96.1 | 67.7 |
| $F_3$ | 1.91 | 5.8 | 10.5 | None | 6.9 | 1.73 | 7.4 | 54.7 |
| $F_4$ | 1.86 | 6.4 | 9.3 | 13.6 | None | 1.69 | 66.1 | 55.8 |
| $F_5$ | 1.85 | 9.8 | 10.8 | None | 14.4 | 1.70 | 55.5 | 52.4 |
| $F_6$ | 1.83 | 3.8 | 9.9 | 5.5 | None | 1.63 | 95.6 | 73.7 |
| $F_7$ | 1.79 | 0.2 | 3.6 | 9.4 | 2.1 | 1.60 | 91.2 | 67.3 |
| $F_8$ | 1.77 | 5.2 | 8.7 | None | 6.3 | | | |
| $F_9$ | 1.74 | 0.2 | 7.8 | 3.2 | 0.9 | | | |
| $F_{10}$ | 1.69 | 3.4 | 10.1 | 2.8 | 0.3 | | | |
| $F_{11}$ | 1.63 | 3.4 | 9.2 | 3.3 | none | | | |

**Table 1:** Summary of characteristic parameters for individual defect families. Description of the different parameters is provided in the main text. To note that, the PSB parameters from Family F8 to F11 are not reported, as the PSBs energy range is not entirely covered by the spectral windows of measurement for these families of defects.

| | Effective and global defect density (/µm²) | |
|---|---|---|
| Flake morphology | Sample A | Sample B |
| Small agglomeration | 0.26 (0.03) | |
| Large agglomeration | 0.53 (0.37) | 0.48 (0.32) |

**Table 2:** Summary of average effective density of defects as a function of hBN flake morphologies. The surface density corresponds to the total number of obsesrved emitters divided by the area covered by hBN flakes, while the global density (value in parenthesis) is calculated by dividing the total number of observed emitters by the overall area being probed through hyperspectral maps.

**Reference:**

1. Koperski, M., Nogajewski, K., Arora, A., Cherkez, V., Mallet, P., Veuillen, J.Y., Marcus, J., Kossacki, P. and Potemski, M., 2015. Single photon emitters in exfoliated WSe$_2$ structures. Nature nanotechnology, 10(6), 503-506.
2. He, Y.M., Clark, G., Schaibley, J.R., He, Y., Chen, M.C., Wei, Y.J., Ding, X., Zhang, Q., Yao, W., Xu, X. and Lu, C.Y., 2015. Single quantum emitters in monolayer semiconductors. Nature nanotechnology, 10(6), 497-502.
3. Tran, T.T., Bray, K., Ford, M.J., Toth, M. and Aharonovich, I., 2016. Quantum emission from hexagonal boron nitride monolayers. Nature nanotechnology, 11(1), 37-41.
4. Palacios-Berraquero, C., Barbone, M., Kara, D. M., Chen, X., Goykhman, I., Yoon, D., Ott, A. K., Beitner, J., Watanabe, K., Taniguchi, T., Ferrari, A. C., Atatüre, M., 2016. Atomically Thin Quantum Light-Emitting Diodes. Nat Commun 7 (1), 12978.
5. Tonndorf, P., Schwarz, S., Kern, J., Niehues, I., Del Pozo-Zamudio, O., Dmitriev, A.I., Bakhtinov, A.P., Borisenko, D.N., Kolesnikov, N.N., Tartakovskii, A.I. and de Vasconcellos, S.M., 2017. Single-photon emitters in GaSe. 2D Materials, 4(2), 021010.
6. Aharonovich, I. and Toth, M., 2017. Quantum emitters in two dimensions. Science, 358(6360), 170-171.
7. Tran, T.T., Wang, D., Xu, Z.Q., Yang, A., Toth, M., Odom, T.W. and Aharonovich, I., 2017. Deterministic coupling of quantum emitters in 2D materials to plasmonic nanocavity arrays. Nano letters, 17(4), 2634-2639.
8. Peyskens, F., Chakraborty, C., Muneeb, M., Van Thourhout, D. and Englund, D., 2019. Integration of single photon emitters in 2D layered materials with a silicon nitride photonic chip. Nature communications, 10(1), 4435.
9. Hötger, A., Klein, J., Barthelmi, K., Sigl, L., Sigger, F., Manner, W., Gyger, S., Florian, M., Lorke, M., Jahnke, F. and Taniguchi, T., 2021. Gate-switchable arrays of quantum light emitters in contacted monolayer MoS$_2$ van der Waals heterodevices. Nano Letters, 21(2), 1040-1046.
10. Su, C., Zhang, F., Kahn, S., Shevitski, B., Jiang, J., Dai, C., Ungar, A., Park, J.H., Watanabe, K., Taniguchi, T. and Kong, J., 2022. Tuning colour centers at a twisted hexagonal boron nitride interface. Nature Materials, 21(8), 896-902.
11. Kianinia, M., Xu, Z.Q., Toth, M. and Aharonovich, I., 2022. Quantum emitters in 2D materials: Emitter engineering, photophysics, and integration in photonic nanostructures. Applied Physics Reviews, 9(1), 011306.
12. Michaelis de Vasconcellos, S., Wigger, D., Wurstbauer, U., Holleitner, A.W., Bratschitsch, R. and Kuhn, T., 2022. Single-Photon Emitters in Layered Van der Waals Materials. physica status solidi (b), 259(4), 2100566.
13. Jungwirth, N.R., Calderon, B., Ji, Y., Spencer, M.G., Flatté, M.E. and Fuchs, G.D., 2016. Temperature dependence of wavelength selectable zero-phonon emission from single defects in hexagonal boron nitride. Nano letters, 16(10), 6052-6057.
14. Martínez, L.J., Pelini, T., Waselowski, V., Maze, J.R., Gil, B., Cassabois, G. and Jacques, V., 2016. Efficient single photon emission from a high-purity hexagonal boron nitride crystal. Physical review B, 94(12), 121405.
15. Tran, T.T., Elbadawi, C., Totonjian, D., Lobo, C.J., Grosso, G., Moon, H., Englund, D.R., Ford, M.J., Aharonovich, I. and Toth, M., 2016. Robust multicolor single photon emission from point defects in hexagonal boron nitride. ACS nano, 10(8), 7331-7338.

16. Exarhos, A.L., Hopper, D.A., Grote, R.R., Alkauskas, A. and Bassett, L.C., 2017. Optical signatures of quantum emitters in suspended hexagonal boron nitride. ACS nano, 11(3), 3328-3336.
17. Li, X., Shepard, G.D., Cupo, A., Camporeale, N., Shayan, K., Luo, Y., Meunier, V. and Strauf, S., 2017. Nonmagnetic quantum emitters in boron nitride with ultranarrow and sideband-free emission spectra. ACS nano, 11(7), 6652-6660.
18. Jungwirth, N.R. and Fuchs, G.D., 2017. Optical absorption and emission mechanisms of single defects in hexagonal boron nitride. Physical review letters, 119(5), 057401.
19. Grosso, G., Moon, H., Lienhard, B., Ali, S., Efetov, D.K., Furchi, M.M., Jarillo-Herrero, P., Ford, M.J., Aharonovich, I. and Englund, D., 2017. Tunable and high-purity room temperature single-photon emission from atomic defects in hexagonal boron nitride. Nature communications, 8(1), 1-8.
20. Museur, L., Feldbach, E. and Kanaev, A., 2008. Defect-related photoluminescence of hexagonal boron nitride. Physical review B, 78(15), 155204.
21. Bourrellier, R., Meuret, S., Tararan, A., Stéphan, O., Kociak, M., Tizei, L.H. and Zobelli, A., 2016. Bright UV single photon emission at point defects in h-BN. Nano letters, 16(7), 4317-4321.
22. Vuong, T.Q.P., Cassabois, G., Valvin, P., Ouerghi, A., Chassagneux, Y., Voisin, C. and Gil, B., 2016. Phonon-photon mapping in a color center in hexagonal boron nitride. Physical review letters, 117(9), 097402.
23. Camphausen, R., Marini, L., Tawfik, S.A., Tran, T.T., Ford, M.J. and Palomba, S., 2020. Observation of near-infrared sub-Poissonian photon emission in hexagonal boron nitride at room temperature. APL Photonics, 5(7), 076103.
24. Hoese, M., Reddy, P., Dietrich, A., Koch, M.K., Fehler, K.G., Doherty, M.W. and Kubanek, A., 2020. Mechanical decoupling of quantum emitters in hexagonal boron nitride from low-energy phonon modes. Science advances, 6(40), eaba6038.
25. Konthasinghe, K., Chakraborty, C., Mathur, N., Qiu, L., Mukherjee, A., Fuchs, G.D. and Vamivakas, A.N., 2019. Rabi oscillations and resonance fluorescence from a single hexagonal boron nitride quantum emitter. Optica, 6(5), 542-548.
26. Preuss, J.A., Groll, D., Schmidt, R., Hahn, T., Machnikowski, P., Bratschitsch, R., Kuhn, T., De Vasconcellos, S.M. and Wigger, D., 2022. Resonant and phonon-assisted ultrafast coherent control of a single hBN color center. Optica, 9(5), 522-531.
27. Gottscholl, A., Kianinia, M., Soltamov, V., Orlinskii, S., Mamin, G., Bradac, C., Kasper, C., Krambrock, K., Sperlich, A., Toth, M. and Aharonovich, I., 2020. Initialization and read-out of intrinsic spin defects in a van der Waals crystal at room temperature. Nature materials, 19(5), 540-545.
28. Stern, H.L., Gu, Q., Jarman, J., Eizagirre Barker, S., Mendelson, N., Chugh, D., Schott, S., Tan, H.H., Sirringhaus, H., Aharonovich, I. and Atatüre, M., 2022. Room-temperature optically detected magnetic resonance of single defects in hexagonal boron nitride. Nature communications, 13(1), 618.
29. Haykal, A., Tanos, R., Minotto, N., Durand, A., Fabre, F., Li, J., Edgar, J.H., Ivady, V., Gali, A., Michel, T. and Dréau, A., 2022. Decoherence of VB− spin defects in monoisotopic hexagonal boron nitride. Nature Communications, 13(1), 4347.
30. Weston, L., Wickramaratne, D., Mackoit, M., Alkauskas, A. and Van de Walle, C.G., 2018. Native point defects and impurities in hexagonal boron nitride. Physical Review B, 97(21), 214104.

31. Sajid, A., Reimers, J.R. and Ford, M.J., 2018. Defect states in hexagonal boron nitride: Assignments of observed properties and prediction of properties relevant to quantum computation. Physical Review B, 97(6), 064101.
32. Sajid, A., Ford, M.J. and Reimers, J.R., 2020. Single-photon emitters in hexagonal boron nitride: a review of progress. Reports on Progress in Physics, 83(4), p.044501.
33. Cholsuk, C., Suwanna, S. and Vogl, T., 2022. Tailoring the Emission Wavelength of Color Centers in Hexagonal Boron Nitride for Quantum Applications. Nanomaterials, 12(14), 2427.
34. Tawfik, S.A., Ali, S., Fronzi, M., Kianinia, M., Tran, T.T., Stampfl, C., Aharonovich, I., Toth, M. and Ford, M.J., 2017. First-principles investigation of quantum emission from hBN defects. Nanoscale, 9(36), 13575-13582.
35. Sajid, A. and Thygesen, K.S., 2020. VNCB defect as source of single photon emission from hexagonal boron nitride. 2D Materials, 7(3), 031007.
36. Mendelson, N., Chugh, D., Reimers, J.R., Cheng, T.S., Gottscholl, A., Long, H., Mellor, C.J., Zettl, A., Dyakonov, V., Beton, P.H. and Novikov, S.V., 2021. Identifying carbon as the source of visible single-photon emission from hexagonal boron nitride. Nature materials, 20(3), 321-328.
37. Koperski, M., Vaclavkova, D., Watanabe, K., Taniguchi, T., Novoselov, K.S. and Potemski, M., 2020. Midgap radiative centers in carbon-enriched hexagonal boron nitride. Proceedings of the National Academy of Sciences, 117(24), 13214-13219.
38. Auburger, P. and Gali, A., 2021. Towards ab initio identification of paramagnetic substitutional carbon defects in hexagonal boron nitride acting as quantum bits. Physical Review B, 104(7), 075410.
39. Jara, C., Rauch, T., Botti, S., Marques, M.A., Norambuena, A., Coto, R., Castellanos-Águila, J.E., Maze, J.R. and Munoz, F., 2021. First-principles identification of single photon emitters based on carbon clusters in hexagonal boron nitride. The Journal of Physical Chemistry A, 125(6), 1325-1335.
40. Tan, Q., Lai, J.M., Liu, X.L., Guo, D., Xue, Y., Dou, X., Sun, B.Q., Deng, H.X., Tan, P.H., Aharonovich, I. and Gao, W., 2022. Donor–acceptor pair quantum emitters in hexagonal boron nitride. Nano Letters, 22(3), 1331-1337.
41. Turiansky, M.E., Alkauskas, A., Bassett, L.C. and Van de Walle, C.G., 2019. Dangling bonds in hexagonal boron nitride as single-photon emitters. Physical review letters, 123(12), 127401.
42. Abdi, M., Chou, J.P., Gali, A. and Plenio, M.B., 2018. Color centers in hexagonal boron nitride monolayers: a group theory and ab initio analysis. ACS Photonics, 5(5), 1967-1976.
43. Reimers, J.R., Shen, J., Kianinia, M., Bradac, C., Aharonovich, I., Ford, M.J. and Piecuch, P., 2020. Photoluminescence, photophysics, and photochemistry of the V B− defect in hexagonal boron nitride. Physical Review B, 102(14), 144105.
44. Ivády, V., Barcza, G., Thiering, G., Li, S., Hamdi, H., Chou, J.P., Legeza, Ö. and Gali, A., 2020. Ab initio theory of the negatively charged boron vacancy qubit in hexagonal boron nitride. npj Computational Materials, 6(1), 41.
45. Chen, Y. and Quek, S.Y., 2021. Photophysical characteristics of boron vacancy-derived defect centers in hexagonal boron nitride. The Journal of Physical Chemistry C, 125(39), 21791-21802.

46. Li, S. and Gali, A., 2022. Identification of an Oxygen Defect in Hexagonal Boron Nitride. The Journal of Physical Chemistry Letters, 13(41), 9544-9551.
47. Chen, Y., Westerhausen, M.T., Li, C., White, S., Bradac, C., Bendavid, A., Toth, M., Aharonovich, I. and Tran, T.T., 2021. Solvent-Exfoliated Hexagonal Boron Nitride Nanoflakes for Quantum Emitters. ACS Applied Nano Materials, 4(10), 10449-10457.
48. Dietrich, A., Bürk, M., Steiger, E.S., Antoniuk, L., Tran, T.T., Nguyen, M., Aharonovich, I., Jelezko, F. and Kubanek, A., 2018. Observation of Fourier transform limited lines in hexagonal boron nitride. Physical Review B, 98(8), 081414.
49. Mendelson, N., Doherty, M., Toth, M., Aharonovich, I. and Tran, T.T., 2020. Strain-induced modification of the optical characteristics of quantum emitters in hexagonal boron nitride. Advanced Materials, 32(21), 1908316.
50. Bommer, A. and Becher, C., 2019. New insights into nonclassical light emission from defects in multi-layer hexagonal boron nitride. Nanophotonics, 8(11), 2041-2048.
51. Feldman, M.A., Puretzky, A., Lindsay, L., Tucker, E., Briggs, D.P., Evans, P.G., Haglund, R.F. and Lawrie, B.J., 2019. Phonon-induced multicolor correlations in hBN single-photon emitters. Physical Review B, 99(2), 020101.
52. Wigger, D., Schmidt, R., Del Pozo-Zamudio, O., Preuß, J.A., Tonndorf, P., Schneider, R., Steeger, P., Kern, J., Khodaei, Y., Sperling, J. and de Vasconcellos, S.M., 2019. Phonon-assisted emission and absorption of individual color centers in hexagonal boron nitride. 2D Materials, 6(3), 035006.
53. Khatri, P., Luxmoore, I.J. and Ramsay, A.J., 2019. Phonon sidebands of color centers in hexagonal boron nitride. Physical Review B, 100(12), 125305.
54. Kern, G., Kresse, G. and Hafner, J.J.P.R.B., 1999. Ab initio calculation of the lattice dynamics and phase diagram of boron nitride. Physical Review B, 59(13), 8551.
55. Wang, L., Pu, Y., Soh, A.K., Shi, Y. and Liu, S., 2016. Layers dependent dielectric properties of two dimensional hexagonal boron nitridenanosheets. AIP Advances, 6(12), 125126.
56. Griffin, A., Harvey, A., Cunningham, B., Scullion, D., Tian, T., Shih, C.J., Gruening, M., Donegan, J.F., Santos, E.J., Backes, C. and Coleman, J.N., 2018. Spectroscopic size and thickness metrics for liquid-exfoliated h-BN. Chemistry of Materials, 30(6), 1998-2005.
57. Hayee, F., Yu, L., Zhang, J.L., Ciccarino, C.J., Nguyen, M., Marshall, A.F., Aharonovich, I., Vučković, J., Narang, P., Heinz, T.F. and Dionne, J.A., 2020. Revealing multiple classes of stable quantum emitters in hexagonal boron nitride with correlated optical and electron microscopy. Nature materials, 19(5), 534-539.
58. Fahey, P.M., Griffin, P.B. and Plummer, J.D., 1989. Point defects and dopant diffusion in silicon. Reviews of modern physics, 61(2), 289.
59. Bourgoin, J., Lannoo, M., 1983. Point Defects in Semiconductors II: Experimental Aspects; Springer Berlin Heidelberg: Berlin, Heidelberg.
60. Hao, T., Ahmed, T., Mou, R.J., Xu, J., Brown, S. and Hossain, Z.M., 2020. Critical inter-defect distance that modulates strength and toughness in defective 2D sp2-lattice. Journal of Applied Physics, 127(20), 204301.
61. Wang, D.S., Ciccarino, C.J., Flick, J. and Narang, P., 2021. Hybridized defects in solid-state materials as artificial molecules. ACS nano, 15(3), 5240-5248.
62. Chen, Y., Li, C., White, S., Nonahal, M., Xu, Z.Q., Watanabe, K., Taniguchi, T., Toth, M., Tran, T.T. and Aharonovich, I., 2021. Generation of high-density quantum emitters in high-quality, exfoliated hexagonal boron nitride. ACS Applied Materials & Interfaces, 13(39), 47283-47292.
63. Vogl, T., Doherty, M. W.; Buchler, B. C.; Lu, Y.; Lam, P. K. 2019. Atomic Localization of Quantum Emitters in Multilayer Hexagonal Boron Nitride. Nanoscale, 11 (30), 14362–14371.
64. Dorn, R.W., Ryan, M.J., Kim, T.H., Goh, T.W., Venkatesh, A., Heintz, P.M., Zhou, L., Huang, W. and Rossini, A.J., 2020. Identifying the molecular edge termination of exfoliated hexagonal boron nitride nanosheets with solid-state NMR spectroscopy and plane-wave DFT calculations. Chemistry of Materials, 32(7), 3109-3121.
65. Sajid, A., Thygesen, K.S., Reimers, J.R. and Ford, M.J., 2020. Edge effects on optically detected magnetic resonance of vacancy defects in hexagonal boron nitride. Communication Physics, 3 (1), 153
66. Provencher, S, 1982. CONTIN: A general purpose constrained regularization program for inverting noisy linear algebraic and integral equations. Computer Physics Communications, 27 (3), 229-249